\documentclass[prl,twocolumn,amsmath,amssymb,superscriptaddress]{revtex4}

\usepackage{graphicx}
\usepackage{hyperref}

\def\const{\mbox{const}}
\def\Re{\mathop{\rm Re}}

\begin{document}

\smallskip

\title{Fermionic full counting statistics with smooth boundaries:\\
from discrete particles to bosonization}

\author{Dmitri~A.~Ivanov}
\affiliation{Institute for Theoretical Physics, ETH Z\"urich,
8093 Z\"urich, Switzerland}
\affiliation{Institute for Theoretical Physics, University of Z\"urich, 
8057 Z\"urich, Switzerland}

\author{Ivan~P.~Levkivskyi}
\affiliation{Institute for Theoretical Physics, ETH Z\"urich,
8093 Z\"urich, Switzerland}
\affiliation{Bogolyubov Institute for Theoretical Physics, 
14-b Metrolohichna Street, Kyiv 03680, Ukraine}

\begin{abstract}
We revisit the problem of full counting statistics of particles on
a segment of a one-dimensional gas of free fermions. Using a combination
of analytical and numerical methods, we study the crossover between the
counting of discrete particles and of the continuous particle density
as a function of smoothing in the counting procedure. In the discrete-particle
limit, the result is given by the Fisher--Hartwig expansion for Toeplitz
determinants, while in the continuous limit we recover the
bosonization results. This example of full counting statistics with
smoothing is also related to orthogonality
catastrophe, Fermi-edge singularity and non-equilibrium bosonization.
\end{abstract}

\date{July 24, 2015}

\maketitle

%%%%%%%%%%%%%%%%%%%%%%%%%
\paragraph{Introduction.---}
Full counting statistics (FCS), in theoretical-condensed-matter context, refers to a class
of problems involving the probability distribution of a quantum observable
(usually the number of electrons found in a certain region of space or transported
through the system over a certain time) with a
particular focus on quantum behavior.
Examples of FCS problems are the anti-bunching of
electrons in a one-dimensional conductor due to their fermionic statistics \cite{levitov:93,levitov:96}
and a single-electron emitter, first proposed theoretically \cite{ivanov:97}
and recently realized experimentally \cite{dubois:13}.

The simplest FCS problems assume non-interacting fermions (electrons), so that
the resulting counting statistics can be expressed in terms of a determinant
taking into account antisymmetrization of the relevant multi-particle processes \cite{levitov:93}.
An alternative approach is based on the bosonization technique (using the equivalence
between bosons and fermions in one dimension) \cite{bosonization}. Bosonization methods for FCS
can be extended to include interactions, but they usually do not fully take
into account the discreteness of particles \cite{levitov:96, levkivskyi:09, gutman:10, levkivskyi:12}.

It is therefore important to understand limitations of the bosonization approach
to FCS and its connection to the exact calculation with discrete fermions. In this Letter,
we address this problem by studying FCS in a one-dimensional free-fermion model, where
an uncertainty (smoothing) in the counting procedure is introduced, so that the particle
number is no longer quantized. In this model, we can study in full detail a
crossover between the discrete-particle and bosonization results as the uncertainty 
increases. While the details of this crossover depend on the profile of the introduced
uncertainty, the qualitative description of the crossover is found to be universal:
FCS respects the discreteness of particles if the uncertainty region is much
narrower than the (average) inter-particle distance, but crosses over to the bosonization result 
when the uncertainty region is much wider than the inter-particle distance.

Mathematically, this problem amounts to studying an evolution of the asymptotic behavior 
of a Toeplitz (or, more precisely, Wiener--Hopf) determinant with a 
Fisher--Hartwig singularity \cite{fisher:68,basor:83}
as the singularity is smoothed
in a certain way. In the case of a sharp singularity, the corresponding determinant
is given by a double asymptotic series of a Fisher--Hartwig type fully respecting 
the particle discreteness \cite{kitanine-kozlowski:08:09,ivanov-abanov-cheianov:13}. 
As the singularity is smoothed, the secondary branches of this expansion get suppressed, 
and the remaining leading branch reproduces the bosonization result.

%In the conclusion of this Letter, we also comment on possible implications of our findings to
%the problem of the Fermi-edge singularity (FES) and other related topics.

\begin{figure}
\centerline{\includegraphics[width=.4\textwidth]{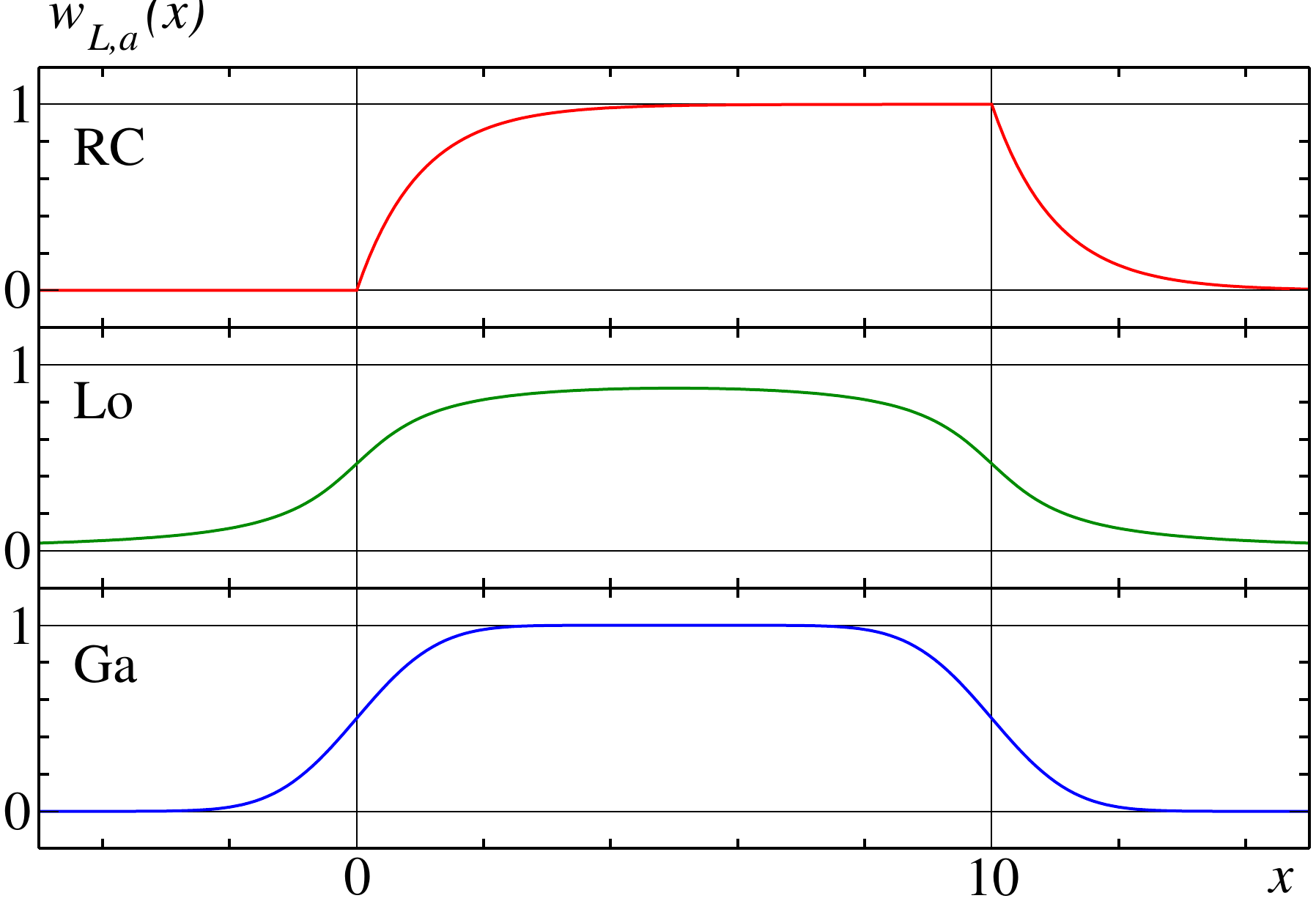}}
\caption{The weight functions $w_{L,a}(x)$ for $L=10$ and $a=1$ 
for the three models of smoothing.}
\label{fig-1}
\end{figure}

\paragraph{Model.---}

We consider the one-dimensional model of spinless free fermions on a line
(both coordinate and momentum are continuous) at zero temperature. 
The system is in its ground state, which is the Slater 
determinant of plane waves characterized by the Fermi wave vector $k_F$: 
the states with the wave vectors smaller than $k_F$ (in absolute value) 
are occupied, and the states with the wave vectors larger than $k_F$ are
empty. We introduce the FCS generating function as \cite{korepin:93,levitov:93}
\begin{equation}
\chi_{L,a}(\kappa)=\left\langle \exp \left(2\pi i \kappa Q_{L,a} \right)\right\rangle\, ,
\label{chi}
\end{equation}
where the average is taken over the ground state. The particle-counting
operator $Q_{L,a}$ is defined as
\begin{equation}
Q_{L,a}=\int_{-\infty}^{\infty} \Psi^\dagger(x) \Psi(x) w_{L,a}(x) \, dx\, ,
\end{equation}
where $\Psi^\dagger(x)$ and $\Psi(x)$ are the fermionic creation and annihilation
operators and $w_{L,a}(x)$ is the weight function for the particle counting.
The weight function depends on the two parameters: the length of the interval $L$ 
and the smoothing length scale $a$, Fig.~\ref{fig-1}. We always assume $L\gg a$.

In the original formulation of the FCS problem, the smoothing is absent
($a=0$), and the weight function is the characteristic function of a
line segment,
\begin{equation}
w_{L,0}(x)=\theta(x)\theta(L-x)=\begin{cases}
0 & \text{$x<0$ or $x>L$,} \\
1 & \text{$0<x<L$.}
\end{cases}
\end{equation}
In this case, $Q_{L,a}$ takes integer values, and
the generating function (\ref{chi}) is periodic in the
counting variable, $\chi_{L,0}(\kappa+1)=\chi_{L,0}(\kappa)$. It can be expressed
as a Toeplitz determinant (see Appendix), which has been subject
to extensive studies \cite{basor:94,ehrhardt:01,calabrese:10,gutman:11,deift:11,krasovsky:11}. 
The periodicity of $\chi_{L,0}(\kappa)$ is reflected in a Fisher--Hartwig-type asymptotic
series (conjectured and numerically verified to a very high order)
\cite{kitanine-kozlowski:08:09,abanov:11,ivanov-abanov-cheianov:13,ivanov-abanov:13}:
\begin{subequations}
\begin{equation}
\chi_{L,0}(\kappa)=\sum_{j=-\infty}^{\infty} \tilde\chi_{L,0}(\kappa+j)\ ,
\label{expansion-1}
\end{equation}
\begin{multline}
\tilde\chi_{L,0} (\kappa) = \exp \Big[
2 \pi i \kappa N_L - 2 \kappa^2 \ln (2 \pi N_L) \\
 + {\tilde C}(\kappa)
+\sum_{n=1}^{\infty} f_n(\kappa)\, (i N_L)^{-n} \Big]\, .
\label{expansion-2}
\end{multline}
\label{expantot}
\end{subequations}
The coefficients in this double expansion can be explicitly calculated: 
${\tilde C}(\kappa)$ is expressed in terms of Barnes G functions as
${\tilde C}(\kappa) = 2 \ln |G(1+\kappa) G(1-\kappa)|$,
and $f_n(\kappa)$ are polynomials in $\kappa$, which can be
computed iteratively, order by order \cite{ivanov-abanov-cheianov:13}
(the same coefficients are denoted $P_{nn}(\kappa)$ in \cite{ivanov-abanov:13}).
$\chi_{L,0}(\kappa)$ depends only on one parameter: the average particle
number on the segment,
\begin{equation}
N_L=k_F L/\pi \, .
\label{N-L}
\end{equation}

\begin{figure}[h!]
\centerline{\includegraphics[width=.45\textwidth]{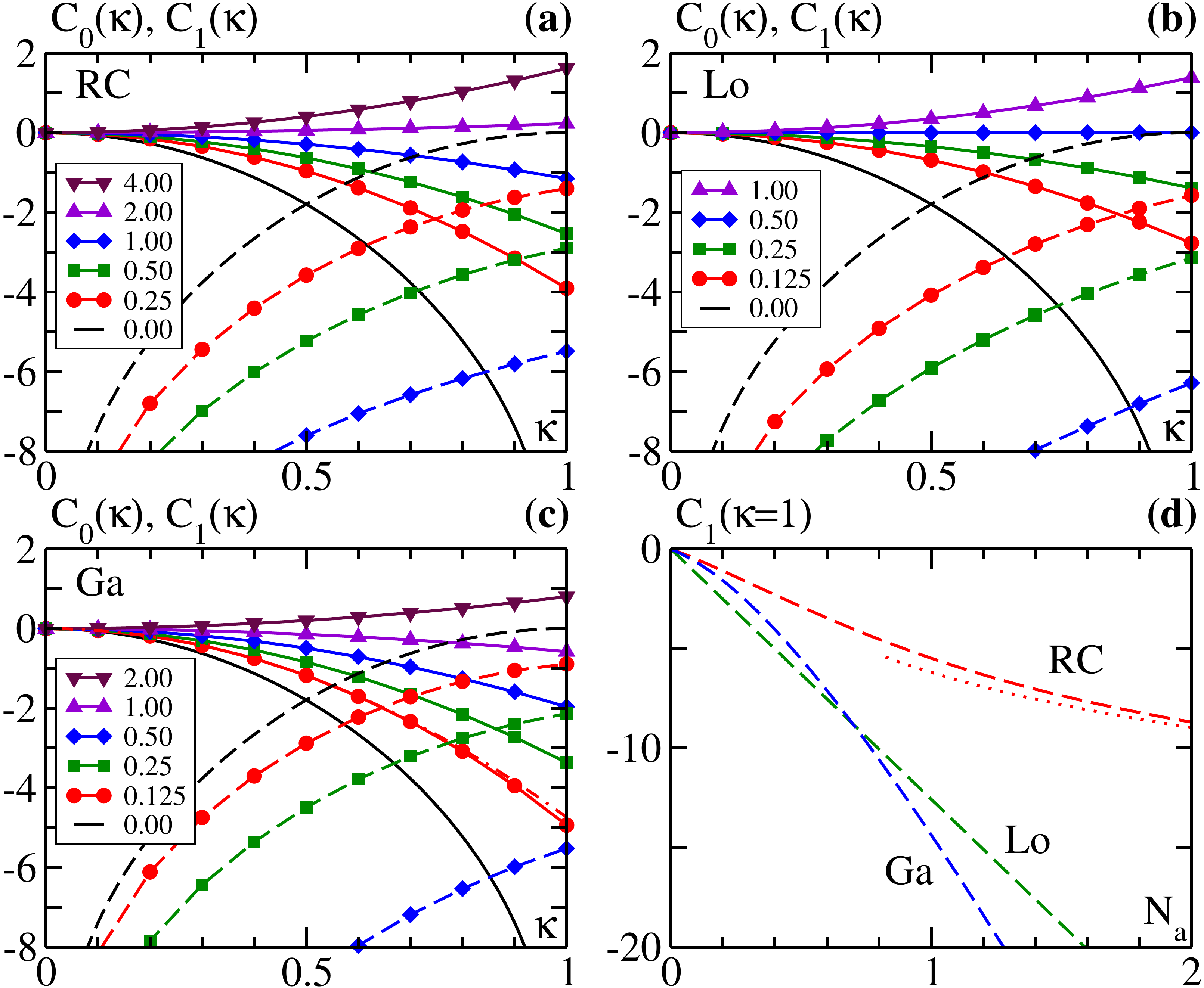}}
\caption{{\bf (a)--(c)}: Coefficients  $C_0(\kappa,a)$ and $C_1(\kappa,a)$
as a function of $\kappa$ for several values of $N_a$, as computed from
numerical fits.
$C_0(\kappa,a)$ are shown by solid lines, $C_1(\kappa,a)$ by dashed lines.
Different symbols correspond to different values of $N_a$ as shown in the legends.
In panel (c), the dash-dotted line is the bosonization result (\ref{bosonization-asymptotics})
at $N_a=0.125$. 
{\bf (d)}: Coefficient  $C_1(\kappa{=}1,a)$ as a function of $N_a$ for the three models. 
The plot for Lo model is the exact analytical result (\ref{C1-Lo-exact}). 
The dotted line is the leading asymptotics (\ref{C1-RC-asympt2}) for RC model.
}
\label{fig-2}
\end{figure}

On the other hand, a bosonization approach proposed for the
generating function $\chi_{L,a}(\kappa)$ \cite{levitov:96}
leads to
\begin{equation}
\chi^{\rm bos}_{L,a}(\kappa)=\exp \Big(
2 \pi i \kappa N_L - 2 \kappa^2 [ \ln (N_L/N_a) + \const ] \Big)\, ,
\label{bosonization-general}
\end{equation}
where we have defined the average number of particles in the smoothing region
\begin{equation}
N_a= k_F a/\pi\, ,
\end{equation}
and the constant at the logarithm depends on the details of the smoothing.
The bosonization expression (\ref{bosonization-general}) corresponds to 
the leading terms in the leading branch ($j=0$)
of the double expansion (\ref{expantot}). We
expect (and confirm it the present work) that this result applies 
provided the smoothing $a$ is sufficiently large. It is not periodic
in $\kappa$, which is consistent with a continuous spectrum
of the operator $Q_{L,a}$ at a finite $a$.

The goal of the present study is to examine in detail the crossover from
the discrete particle counting given by the expansion 
(\ref{expantot}) to the bosonization result
(\ref{bosonization-general}) as a function of the smoothing parameter $a$.
Some of the results are derived analytically, while a more detailed
picture of the crossover is obtained numerically. Smoothing 
is introduced as a convolution
\begin{equation}
w_{L,a}(x)=\int_{-\infty}^{\infty} \frac{dx'}{a}\; \theta(x)\theta(L-x)\; g(x'/a) \, ,
\label{smoothing}
\end{equation}
where $g(\tilde{x})$ is a smoothing function for the scaled coordinate $\tilde{x}=x/a$.
We always assume the normalization $\int d\tilde{x}\; g(\tilde{x})=1$.
To be specific, we consider three models of smoothing corresponding
to the three choices of the smoothing function: RC 
(this type of smoothing may be relevant for particle-counting in time,
where the counter has a RC-type response function),
Lorentzian (Lo) and Gaussian (Ga):
\begin{equation}
g(\tilde{x})=\begin{cases}
\theta(\tilde{x})\exp(-\tilde{x}) & \text{RC,} \\
[\pi(\tilde{x}^2+1)]^{-1} & \text{Lo,} \\
\exp(-\tilde{x}^2/2)/\sqrt{2\pi} & \text{Ga}
\end{cases}
\end{equation}
(see Fig.~\ref{fig-1}). In any of these models, the FCS generating function $\chi_{L,a}(\kappa)$
depends on $L$ and $a$ via the two dimensionless parameters $N_L$ and $N_a$.

\paragraph{Results.---}

We conjecture (and support this conjecture by analytical and numerical results)
that, at a finite $a$, the FCS generating function admits the same structure of the
expansion (\ref{expantot}) as at $a=0$, 
except that the expansion coefficients
are not periodic in the branch index $j$ and depend on the smoothing parameter $a$.
If we limit the range of $\kappa$ to the interval $[0,1]$, we can write
the leading terms of the expansion as
\begin{multline}
\chi_{L,a}(\kappa) \approx \exp \Big[
2 \pi i \kappa N_L - 2 \kappa^2 \ln N_L +C_0(\kappa,a) \Big] \\
+ \exp \Big[
2 \pi i (\kappa-1) N_L - 2 (\kappa-1)^2 \ln N_L +C_1(\kappa,a) \Big] \, .
\label{expansion-a}
\end{multline}
At $a=0$, in agreement with the full expansion (\ref{expantot}),
the coefficients $C_0(\kappa,a)$ and $C_1(\kappa,a)$ are given by
\begin{multline}
C_0(\kappa,0)=C_1(1-\kappa,0)\\
=2 \ln \left| G(1+\kappa) G(1-\kappa) \right| -2\kappa^2\ln (2\pi)\, .
\end{multline}
We expect that, as $a$ increases, $C_0(\kappa,0)$ also increases to reproduce the
bosonization result (\ref{bosonization-general}), while $C_1(\kappa,0)$ decreases
to suppress the second term in the expansion (\ref{expansion-a}).

This behavior is indeed confirmed by numerical studies. We report the
details of our numerics in the Appendix, an here we only plot in
Fig.~\ref{fig-2}a-c the results of the numerical fits for the coefficients 
$C_0(\kappa,a)$ and $C_1(\kappa,a)$ for the three models of the smoothing.
Figure \ref{fig-2}d shows more detailed results for the coefficient $C_1(1,a)$ representing
the suppression of the secondary Fisher--Hartwig branch at $\kappa=1$. This coefficient
was, in fact, computed as a doubled contribution from a single ``phase slip'' 
(at the beginning or at the end of the interval) in the weight function $w_{L,a}(x)$
(note that at $\kappa=1$ the values $w_{L,a}(x)=0$ and $w_{L,a}(x)=1$ are equivalent,
see Appendix).
An agreement between the results computed with this method (at $\kappa=1$)
and the fitting procedure for the full function $\chi_{L,a}(\kappa)$
serves as an independent check of our numerical scheme.

As expected, while the actual values of the coefficients 
$C_0(\kappa,a)$ and $C_1(\kappa,a)$ depend on the chosen smoothing model,
qualitatively the crossover between the discrete particle counting and a continuous
bosonization description occurs at $N_a\sim 1$ in all the three models.

Furthermore, some of the numerical results presented above can be verified by analytical
means. First, the constant in the bosonization formula (\ref{bosonization-general}) can
be computed from the corresponding Toeplitz determinant using the strong Szeg\H{o} theorem 
\cite{szego:52,boettcher:90}.
This results in the asymptotic behavior for $C_0(\kappa,a)$:
\begin{equation}
C_0(\kappa, a\to\infty) = 2\kappa^2(\ln N_a - \Upsilon)\, ,
\label{bosonization-asymptotics}
\end{equation}
where
\begin{equation}
\Upsilon = \gamma + \lim_{\varepsilon \to 0} \left[ \int_\varepsilon^\infty \frac{dk}{k} g_k \; g_{-k} 
+ \ln \varepsilon\right] \, ,
\label{upsilon-def}
\end{equation}
in the case of a general smoothing model (\ref{smoothing}).  Here $\gamma=0.5772\ldots$ is the
Euler--Mascheroni constant and $g_k$ are the Fourier components
of the smoothing function,
\begin{equation}
g_k=\int_{-\infty}^{\infty} d\tilde{x}\, e^{-i k\tilde{x}} g(\tilde{x})\, .
\end{equation}
For the three models considered in our paper, a calculation gives
\begin{equation}
\Upsilon_{\rm RC}=\gamma\, , \qquad \Upsilon_{\rm Lo}=-\ln 2\, , \qquad
\Upsilon_{\rm Ga}=\gamma/2\, .
\label{upsilon-values}
\end{equation}
These analytical results perfectly agree with our numerical fits: for the RC and Lorentzian smoothing,
the computed values of $C_0(\kappa,a)$ for $N_a\ge 0.25$ and $N_a\ge 0.125$, respectively, are 
indistinguishable in the plots of Fig.~\ref{fig-2}a,b from the bosonization asymptotics (\ref{bosonization-asymptotics}),
(\ref{upsilon-values}). For the Gaussian smoothing,
the difference between $C_0(\kappa,a)$ and the bosonization asymptotics is visible for $N_a=0.125$,
but not for $N_a\ge 0.25$ (Fig.~\ref{fig-2}c).

Second, the asymptotic behavior of $C_1(1,a)$ as $a\to \infty$ can also be calculated analytically
using an asymptotic formula for the Toeplitz determinant with a non-zero winding number \cite{boettcher:06}.
A calculation results in
\begin{multline}
\frac{1}{2}C_1(1,a\to\infty) \approx \Xi -\Upsilon \\ 
+ \Re \ln \int_{-\infty}^\infty\frac{dz}{2\pi} 
\exp\left[2\pi i N_a z -\int_{-\infty}^{\infty} dk\; g_k \frac{1-e^{ikz}}{|k|} \right]\, ,
\label{C1-asympt1}
\end{multline}
where $\Upsilon$ is given by (\ref{upsilon-def}) and $\Xi$ is defined as
\begin{equation}
\Xi=2\gamma + \lim_{\varepsilon\to 0} \left[ 
\int_\varepsilon^\infty \frac{dk}{k} (g_k+g_{-k}) + 2\ln\varepsilon \right]\, .
\label{xi-def}
\end{equation}
An explicit calculation gives
\begin{equation}
\Xi_{\rm RC}=2 \gamma\, , \qquad \Xi_{\rm Lo}=0\, , \qquad
\Xi_{\rm Ga}=\gamma+ \ln 2\, .
\label{xi-values}
\end{equation}
Remarkably, the asymptotics (\ref{C1-asympt1}), while not exactly coinciding with 
$C_1(1,a)$, gives a very good approximation in the whole range of the values of $a$
(in the plot in Fig.~\ref{fig-2}d it would be indistinguishable from the exact
values, see details in Appendix).

The analytic approximation (\ref{C1-asympt1}) also allows us to extract the 
leading asymptotic behavior. For the RC model, we find
\begin{equation}
C_1(1,a\to\infty)_{\rm RC} \approx 2 \gamma - 4 \ln(2 \pi N_a) \, ,
\label{C1-RC-asympt2}
\end{equation}
i.e., the secondary branch of $\chi(1,a)$ decays as $N_a^{-4}$. 
For the Lorenzian smoothing, the Toeplitz determinant for a single ``phase slip''
$w_a(x)=\frac{1}{2\pi i}\ln(\frac{x+ia}{x-ia})$ may be computed exactly by using
the analyticity of the function $\exp[2\pi w_a(x)]$ in one of the half-planes. An
argument in the spirit of \cite{ivanov:97,keeling:06} then leads to the exact result:
\begin{equation}
C_1(1,a)_{\rm Lo} = -4\pi N_a\, .
\label{C1-Lo-exact}
\end{equation}
Finally, for the Gaussian smoothing, the asymptotic behavior is
\begin{equation}
C_1(1,a\to\infty)_{\rm Ga} \sim -4\pi N_a \sqrt{\ln (2 \pi N_a^2)}\, .
\label{C1-Ga}
\end{equation}
However, this expression by itself [unlike the integral formula (\ref{C1-asympt1})]
does not provide a good approximation for $C_1(1,a)$, since the next-order corrections
are not $O(1)$.

Remarkably, even though the coefficient at the secondary branch is reduced with increasing $N_a$,
the dependence on $N_L$ does not change. In particular, for $\kappa>1/2$, the bosonization (first) term
in (\ref{expansion-a}) always decreases with $N_L$ faster than the non-bosonization (second) one. This
implies that the non-bosonization corrections arising from the discreteness of particles should
be visible, in this range of $\kappa$, at sufficiently large $L$ (see Appendix). The corresponding values 
of $L$ depend on the particular choice of smoothing and can be deduced from comparing the two terms 
in (\ref{expansion-a}) using the asymptotics at large $N_a$ derived above. A practical observation of this 
effect may be limited at large $N_a$ if both terms in (\ref{expansion-a}) become too small.

\paragraph{Discussion.---}
In this Letter, we have discussed the crossover from discrete to continuous FCS
in the model of free one-dimensional fermions, as a function of smoothing. This model may serve
as an illustration of a relation between discrete and continuous descriptions
in a wide spectrum of similar problems, including FCS with temporal measurements,
Fermi-edge singularity, orthogonality catastrophe, and non-equilibrium bosonization.

By FCS with temporal measurements we understand a FCS setup where the counting is performed
over a certain time interval, e.g., by opening and closing an electric contact or by
applying a time-dependent voltage pulse \cite{levitov:93,levitov:96}. 
This formulation of FCS used in the
original FCS papers differs from our approach where the measurement is extended
in space instead of time. This difference implies the necessity of regularizing
the contribution of the Fermi sea in temporal FCS: a large current of left-movers is compensated
by a similarly large current of right-movers, so the average charge transfer is determined
by the vicinity of the Fermi level, while the quantum noise depends on the ultraviolet
cutoff defined by the Fermi energy $\varepsilon_F$ (arising from the bottom of the Fermi sea). 
In our spatial formulation, there is no cutoff introduced by the bottom of the Fermi sea, 
and both the total number of particles and its fluctuations depend on the same scale $k_F$. 
As a consequence of this difference in ultraviolet cutoffs, our results cannot be literally 
translated to the case of temporal FCS. However, we expect that the main qualitative conclusion
will hold also in the temporal case: the generating function $\chi(\kappa)$ loses
periodicity in $\kappa$ when the smoothing time scale exceeds the typical
time between particles (i.e., $\varepsilon_F^{-1}$).

Determinants similar to the generating function $\chi(\kappa)$ also appear in problems
related to the orthogonality catastrophe and to the Fermi-edge singularity (FES). In those cases,
the counterparts of the secondary branches of $\chi(\kappa)$ are secondary singularities
(cusps or peaks) in the frequency-dependent response function (the closed-loop contribution in
the FES context). Such secondary cusps and peaks were studied, e.g., in the recent work \cite{knap:12}
(and, in the case of a bound-state contribution, earlier in \cite{combescot:71}).
According to our predictions, such cusps and peaks should be most visible in case
of instant switching of the scattering potential, but get suppressed if
the switching time of the scattering potential exceeds $\varepsilon_F^{-1}$.

We should also remark that most studies of FES involve an artificial regularization
of the Fermi-sea contribution at energy scales smaller than $\varepsilon_F$, so that
secondary singularities in the response function are neglected
\cite{nozieres:69, abanin:05, chernii:14}. Such secondary singularities
(separated from the main peak by $\varepsilon_F$) are 
probably not experimentally relevant in physical metals, where $\varepsilon_F$
is a large energy scale, but may be of interest in other models with FES physics
where $\varepsilon_F$ is, for some reason, small (see, e.g., 
Ref.~\cite{tikhonov:10} for an example from spin-liquid theory).

Finally, we also mention recent works on non-equilibrium bosonization
where determinants similar to ours appear (again, in the temporal form) 
\cite{gutman:10}. Similarly to the FES problem discussed above,
these works assume a regularization of the Fermi-sea contribution, which
is equivalent to neglecting the secondary branches of the Fisher--Hartwig
expansion related to the bottom of the Fermi sea.

\paragraph{Acknowledgment.---}
The authors are grateful to A.~G.~Abanov, E.~Demler, M.~V.~Feigelman, L.~Glazman,
L.~Levitov, A.~Mirlin, and E.~V.~Sukhorukov for helpful discussions. The work of
D.A.I.\ was supported by the Swiss National Foundation
through the NCCR QSIT. I.P.L.\ was supported by Marie Curie Actions COFUND program.

\section{Appendix}

\paragraph{A. Toeplitz determinant for $\chi(\kappa)$.---}
The FCS generating function (\ref{chi}) can be written as a trace
in the multi-particle space and then re-expressed as a determinant
in the single-particle space \cite{levitov:93,levitov:96,klich:03}:
\begin{equation}
\chi_{L,a}(\kappa)=\det\left[ (1-n_F) + n_F\; e^{2\pi i \kappa w} \right]\, ,
\label{determinant-1}
\end{equation}
where $n_F$ is the Fermi occupation number and $w$ is the operator
of multiplication by $w_{L,a}(x)$. The operator $n_F$ is diagonal
in the momentum space, while $w$ is diagonal in the coordinate space.
In this paper, we consider the zero-temperature case, so $n_F$ is a projector
onto the occupied states (below $k_F$).

In the limit of no smoothing ($a=0$), $w$ is also a projector (in the
coordinate space), and the determinant is symmetric with respect
to exchanging coordinate and momentum:
\begin{equation}
\chi_{L,0}(\kappa)=\det\left[ 1- (1- e^{2\pi i \kappa}) n_F w \right]\, .
\end{equation}
This operator is of Toeplitz (or Wiener--Hopf) form in both coordinate
and momentum representations.

After introducing smoothing, the operator in (\ref{determinant-1}) is
no longer Toeplitz in the coordinate representation, but remains Toeplitz
in the momentum representation (at zero temperature), 

\paragraph{B. Toeplitz determinant for $C_1(\kappa{=}1,a)$.---}
At $\kappa=1$, the term with $C_1(\kappa,a)$ in the decomposition
(\ref{expansion-a}) does not oscillate with $L$. It therefore 
arises from the end points of the segment of measurement
($x=0$ and $x=L$). This allows us to calculate $C_1(\kappa{=}1,a)$
directly from a single-step contribution:
\begin{equation}
\frac{1}{2} C_1(\kappa{=}1,a) = \Re \ln \det
\left[ (1-n_F) + n_F e^{2\pi i \tilde{w}} \right]\, ,
\label{det-one-step}
\end{equation}
where $\tilde{w}$ is the operator of multiplication by the single-step
counterpart of $w_{L,a}(x)$:
\begin{equation}
\tilde{w}_a(x) = \int_{-\infty}^{\infty} \frac{dx'}{a}\; 
\theta(x)\; g(x'/a) \, .
\end{equation}
Note that while $\tilde{w}_a(x)$ has different limits at $x\to\pm\infty$,
the exponent $\exp(2\pi i \tilde{w})$ tends to 1 in both limits, and the
determinant (\ref{det-one-step}) is well defined.

\begin{figure}[t]
\centerline{\includegraphics[width=.45\textwidth]{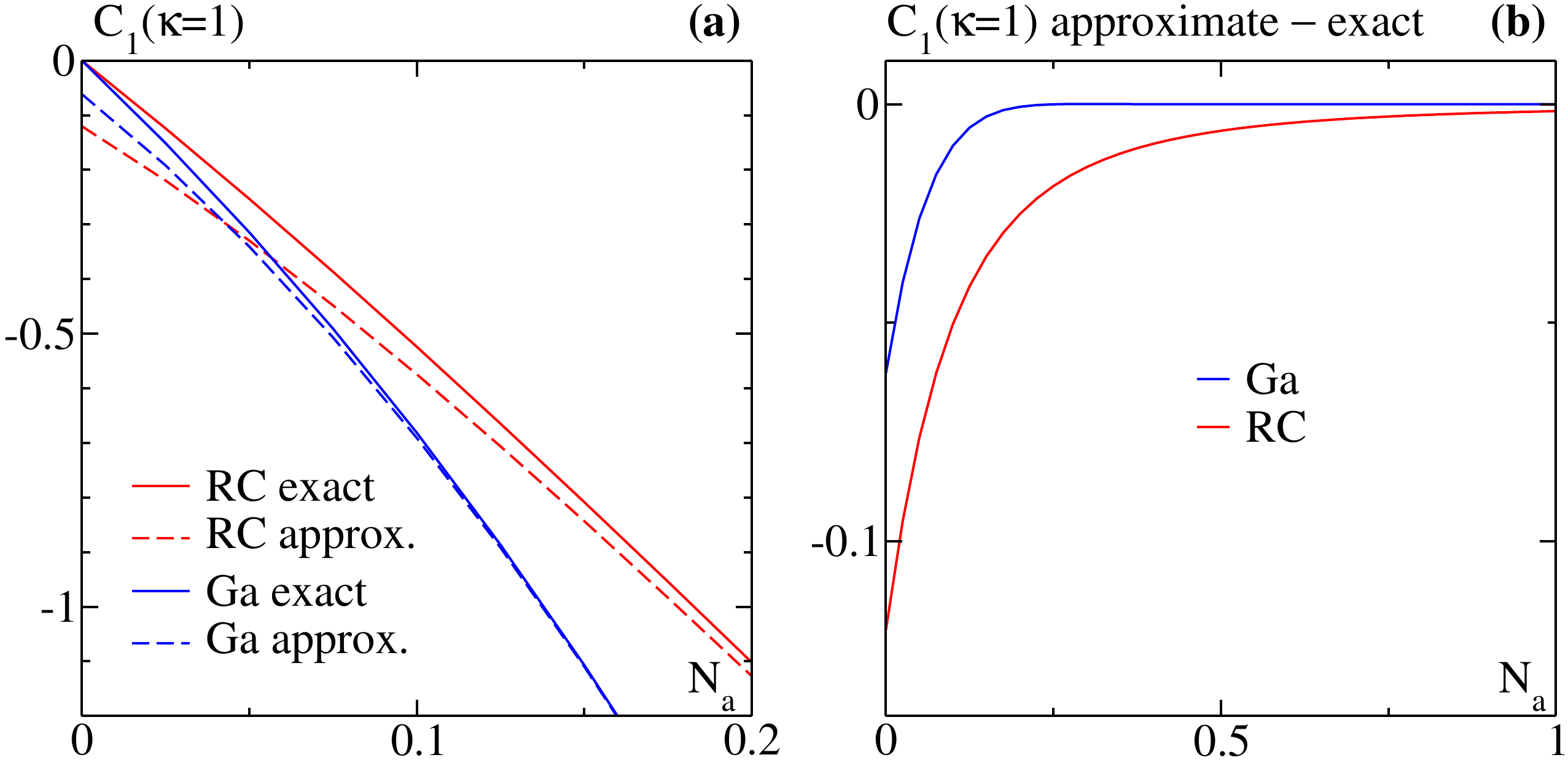}}
\caption{{\bf (a)} the values of $C_1(\kappa{=}1,a)$ for RC and Ga models
(the same as in Fig.~\ref{fig-2}d) as calculated numerically, together with the
analytic approximations given by (\ref{C1-asympt1}). {\bf (b)} the difference between
the approximate and exact values of $C_1(\kappa{=}1,a)$ (upper curve: Ga model, 
lower curve: RC model).}
\label{one_tail_analytic}
\end{figure}

\paragraph{C. Details of the numerical calculation.---}
For numerical calculations of $\chi_{L,a}(\kappa)$ and $C_1(\kappa{=}1,a)$, 
we approximate the integral operators (\ref{determinant-1}) and (\ref{det-one-step})
by finite-dimensional matrices by discretizing the momentum space 
(which is equivalent to considering the system on a circle). In the momentum
space, these matrices are of Toeplitz form. The determinants of those matrices 
are calculated numerically for several matrix sizes and then extrapolated to the
infinite matrix size in order to obtain $\chi_{L,a}(\kappa)$ and $C_1(\kappa{=}1,a)$, 
respectively.

For numerical calculations of the determinants, we use the LAPACK library \cite{LAPACK},
and Fourier transformations necessary for computing the matrix elements were done
with the help of the GNU scientific library \cite{gsl}. In our calculations, determinants
of matrices of the linear size up to 600 were calculated, which allowed us to achieve
good precision for the considered ranges of parameters.
 
The values of $\chi_{L,a}(\chi)$ were then fitted to the expansion (\ref{expansion-a}), 
and the coefficients $C_0(\kappa,a)$ and $C_1(\kappa,a)$ were extracted 
(the fit also included terms up to $L^{-2}$ in the exponents). 
As an additional check of this fitting procedure, we have verified that the values 
of $C_1(\kappa{=}1,a)$ obtained from these fits agree with those calculated 
independently using the one-step method described above.

\begin{figure}[t]
\centerline{\includegraphics[width=.45\textwidth]{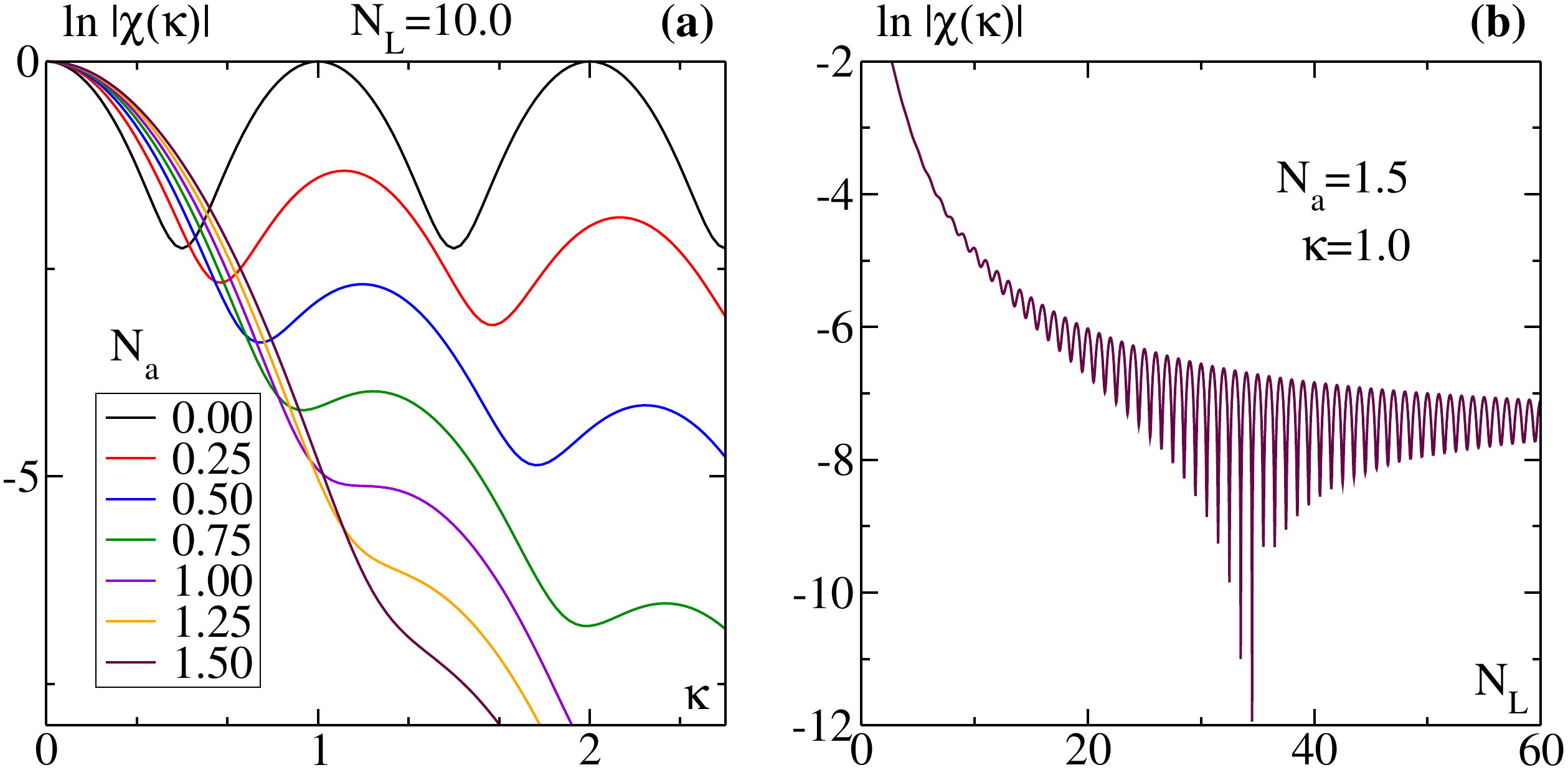}}
\caption{
{\bf (a)} $\ln|\chi_{L,a}(\kappa)|$ as a function of $\kappa$ in the RC model
at $N_L=10$ for selected values of $N_a$. Periodicity in $\kappa$ is lost with
increasing $N_a$. {\bf (b)} $\ln|\chi_{L,a}(\kappa)|$ as a function of $N_L$ in the RC model
at $\kappa=1$ and $N_a=1.5$. For large $N_L$, the second term in 
(\ref{expansion-a}) dominates.}
\label{two-plots}
\end{figure}

\paragraph{D. Analytic approximation for $C_1(\kappa{=}1,a{\to}\infty)$.---}
The approximation (\ref{C1-asympt1}), while derived under the assumption $N_a \gg 1$,
is remarkably accurate even at small $N_a$. In the Lo model, where $C_1(\kappa{=}1,a)$
is given exactly by (\ref{C1-Lo-exact}), the approximation (\ref{C1-asympt1}) reproduces
the same exact result for all $N_a$. In the RC and Ga models, the difference between
(\ref{C1-asympt1}) and $C_1(\kappa{=}1,a)$ tends to zero as $a\to\infty$, but remains
finite at finite $a$. However, this difference is numerically small for these two
models, even at $a=0$. We plot this difference as a function of $N_a$ in 
Fig.~\ref{one_tail_analytic}.

\paragraph{E. Numerical illustration of the crossover.---}
We illustrate the crossover described in the main body of the paper by two plots.
In Fig.~\ref{two-plots}a, we show how the periodicity of $\chi_{L,a}(\kappa)$ in $\kappa$
is gradually lost as $N_a$ increases. Figure \ref{two-plots}b illustrates how the second
term in (\ref{expansion-a}) dominates at large $L$ (in RC model, for $\kappa=1$, this happens
at $L \gg k_F^2a^3$). 

%%%%%%%%%%%%%%%%%%%%%%%%%%%%%%%%%%%%%%%
%%%%%%%%%%%%%%%%%%%%%%%%%%%%%%%%%%%%%%%
%%%%%%%%%%%%%%%%%%%%%%%%%%%%%%%%%%%%%%%

%%%%%%%%%%%%%%%%%%%%%%%%%%%%%%%%%%%%%%%
%%%%%%%%%%%%%%%%%%%%%%%%%%%%%%%%%%%%%%%
%%%%%%%%%%%%%%%%%%%%%%%%%%%%%%%%%%%%%%%

\end{document}